# Electrically controlled emission from singlet and triplet exciton species in atomically thin light emitting diodes

Andrew Y. Joe[1], Luis A. Jauregui[2], Kateryna Pistunova[1], Andrés M. Mier Valdivia[3], Zhengguang Lu[4,5], Dominik S. Wild[1], Giovanni Scuri[1], Kristiaan De Greve[1,6+], Ryan J. Gelly[1], You Zhou[1,6], Jiho Sung[1,6], Andrey Sushko[1], Takashi Taniguchi[7], Kenji Watanabe[8], Dmitry Smirnov[4], Mikhail D. Lukin[1], Hongkun Park[1,6] and Philip Kim[1,3]*

[1] *Department of Physics, Harvard University, Cambridge, Massachusetts 02138, USA*
[2] *Department of Physics and Astronomy, University of California, Irvine, California 92697, USA*
[3] *John A. Paulson School of Engineering and Applied Sciences, Harvard University, Cambridge, Massachusetts 02138, USA*
[4] *National High Magnetic Field Laboratory, Tallahassee, Florida 32310, USA*
[5] *Department of Physics, Florida State University, Tallahassee, Florida 32306, USA*
[6] *Department of Chemistry and Chemical Biology, Harvard University, Cambridge, Massachusetts 02138, USA*
[7] *International Center for Materials Nanoarchitectonics, National Institute for Materials Science, 1-1 Namiki, Tsukuba 305-0044, Japan*
[8] *Research Center for Functional Materials, National Institute for Materials Science, 1-1 Namiki, Tsukuba 305-0044, Japan*
+ *currently at IMEC, 3001 Leuven, Belgium*

*\* To whom correspondence should be addressed: pkim@physics.harvard.edu*

**Excitons are composite bosons that can feature spin singlet and triplet states. In usual semiconductors, without an additional spin-flip mechanism, triplet excitons are extremely inefficient optical emitters. Transition metal dichalcogenides (TMDs), with their large spin-orbit coupling, have been of special interest for valleytronic applications for their coupling of circularly polarized light to excitons with selective valley and spin[1–4]. In atomically thin MoSe$_2$/WSe$_2$ TMD van der Waals (vdW) heterostructures, the unique atomic registry of vdW layers provides a quasi-angular momentum to interlayer excitons[5,6], enabling emission from otherwise dark spin triplet excitons. Here, we report electrically tunable spin singlet and triplet exciton emission from atomically aligned TMD heterostructures. We confirm the spin configurations of the light-emitting excitons employing magnetic fields to measure effective exciton $g$-factors. The interlayer tunneling current across the TMD vdW heterostructure enables the electrical generation of singlet and triplet exciton emission in this atomically thin PN junction. We demonstrate electrically tunability between the singlet and triplet excitons that are generated by charge injection. Atomically thin TMD heterostructure light emitting diodes thus enables a route for optoelectronic devices that can configure spin and valley quantum states independently by controlling the atomic stacking registry.**

Inefficient light emission from triplet excitons is known to be a major bottleneck for many optoelectronic devices, such as e.g. organic LEDs, where a significant portion of randomly generated excitons are in spin-triplet states[7]. The search for highly emissive "bright" triplet excitons generally involves materials with strong spin-orbit coupling (SOC)[8] that also provide such spin-flip mechanisms with altered optical selection rules. Electronic band structure engineering in van der Waals (vdW) heterostructures of transition metal dichalcogenides (TMDs) with strong SOC may produce a unique material platform to realize optoelectronic devices with emissive triplet excitons.

Semiconducting TMDs exhibit extraordinary excitonic effects when reduced to the two-dimensional limit[9–11]. Monolayer TMDs have large exciton binding energies[12] and spin-valley locking[13,14], which can be harnessed for optoelectronic[15,16] and valleytronic[14] applications. When certain monolayer TMDs are stacked together to form heterobilayers such as WSe$_2$/MoSe$_2$[1–4,17–21], MoSe$_2$/MoS$_2$[22], or WS$_2$/MoS$_2$[23–25], energetically favorable interlayer excitons (IEs) can form across the atomically sharp interfaces owing to their type-II band alignment[26] and ultrafast charge transfer[22–24] between the layers. The resulting IEs have long lifetimes[17,19,20], a permanent out-of-plane dipole moment[3,19], and modified optical selection rules[2–6] due to the electrons and holes residing in separate layers. When the heterostructures are electron or hole doped, the IEs bind with free carriers to form charged interlayer excitons (CIEs)[19]. Furthermore, the IEs are predicted to have modified selection rules allowing emission from both singlet (electron and hole with opposite spin) and triplet (electron and hole with same spin) excitons with opposite polarization properties[5,6], which provides a material platform with two long-lived, electrically-tunable IE states.

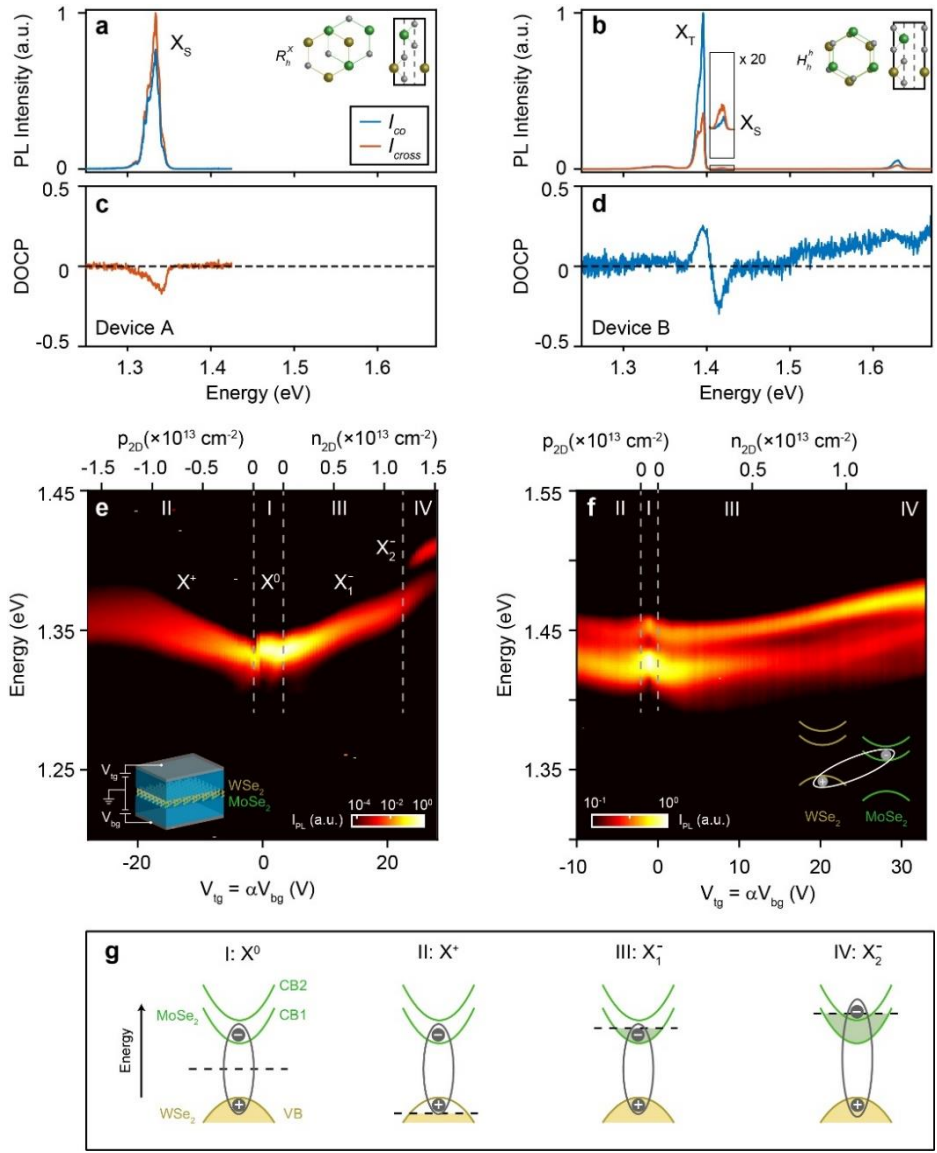

**Figure 1 | Photoluminescence of interlayer excitons in MoSe$_2$/WSe$_2$ heterostructures. a, b,** Polarization resolved photoluminescence (PL) from devices A and B, respectively. Excitation power, P = 0.5 µW and 100 µW, respectively, are used to compare all existing exciton species in the undoped regime. Inset: Lowest energy stacking configuration for 0- and 60-degree heterostructures. **c, d,** Degree of circular polarization (DOCP) extracted from a-b. **e, f,** PL vs. $V_{tg} = \alpha\, V_{bg}$, where $\alpha$ is based on the top and bottom $h$-BN thicknesses, for devices A and B, respectively. Left inset: device schematic and direction of applied gate voltages. Right inset: band schematic of an interlayer exciton between the conduction band of MoSe$_2$ and valence band of WSe$_2$. **g,** Reduced band diagrams of the MoSe$_2$/WSe$_2$ heterostructure showing the upper (CB2) and lower (CB1) conduction bands of MoSe$_2$ and the valence band of WSe$_2$ (VB). The exciton and Fermi energy (black dashed line) is drawn for each regime marked in **(e, f)**. The green (MoSe$_2$) and yellow (WSe$_2$) shaded areas indicate filled electron bands.

Since the optical selection rules differ for different atomic registries of the TMD heterostructure, the IE emission spectra, even qualitatively, vary between studies without consistent observation of both singlet and triplet states or specifications of the heterostructure stacking configuration[2–4,27,28]. Theoretical studies have predicted selection rules for TMD heterostructures based on the stacking orientation, either 0-degree aligned or 60-degree aligned, and atomic registry, which provide a quasi-angular momentum to brighten singlet and triplet optical transitions[5,6]. Since both theoretical[6] and experimental studies[2] claim a dominant atomic registry exists for either the 0- and 60-degree cases, we focus on these specific twisting angles. In this work, we demonstrate electrostatic doping and bias-controlled spin singlet and triplet exciton emission in a 0- and 60-degree MoSe$_2$/WSe$_2$ heterostructure.

Our experiments employ $h$-BN encapsulated WSe$_2$/MoSe$_2$ devices with top and bottom gates, and electrically transparent contacts (Fig. 1e left inset), as described in our previous work[19]. We use a dual-gating scheme where the top-gate voltage ($V_{tg}$) and the back-gate voltage ($V_{bg}$) have the same polarity, achieving higher carrier densities than in previous IE studies[3,19,20] (details in Supplementary Section 1). Furthermore, separate electrical contacts made for MoSe$_2$ and WSe$_2$ layers allows operating the device as an atomically thin PN diode where the current can flow across the vdW interface. Below we focus on two representative devices with 0-degree (device A) and 60-degree (device B) stacking orientations.

Figs. 1a-b show a comparison of the photoluminescence (PL) spectrum for device A and B at neutral doping with circularly co- and cross-polarized exciton emission ($I_{co}$ and $I_{cross}$, respectively). In device A, we observe only a single peak at ~ 1.34 eV. For device B, there are two peaks, one at ~ 1.39 eV and the other at ~ 1.41 eV, with a separation of ~ 25 meV. As shown in Figs. 1c-d, the degree of circular polarization (DOCP), computed from $\frac{I_{co}-I_{cross}}{I_{co}+I_{cross}}$, of the lower energy peak at 1.39 eV in device B is positive, suggesting the chirality of the light emitted remains unchanged, unlike the higher energy peak in device B and the peak in device A. The observation of two peaks with opposite DOCP in device B is consistent with previous experimental results for 60-degree aligned heterostructures[27,28] and their selection rules[5]. Thus, we tentatively identify the lower and higher energy peaks as triplet ($X_T$) and singlet excitons ($X_S$), respectively, while the emission energy and DOCP for device A are consistent with a 0-degree heterostructure[2].

The PL spectrum can further be modified by applying gate voltages of the devices. Fig. 1e-f show the PL spectrum from devices A and B as a function of $V_{tg} = \alpha\, V_{bg}$, where $\alpha = 0.617$ or $1.4$ for the two devices (based on each device's $h$-BN thicknesses), respectively. In this gate voltage configuration, we can maximize our achievable 2D carrier density $n_{2D}$. We identify four distinct gate regions, marked by I-IV, from the electrostatic doping of the heterostructure (Fig. 1g). We verify the doping of the layers by measuring the intralayer exciton absorption spectra as a function of the gate voltage (Supplementary Fig. 3).

The gate dependent PL shows strong atomic stacking registry dependence. For device A (Fig. 1e), only neutral interlayer excitons labeled as $X^0$ appear in region I. In region II (III), the Fermi energy crosses the valence band of WSe$_2$ (lower conduction band (CB1) of MoSe$_2$) and we begin to p-dope (n-dope) the heterostructure forming CIEs, $X^+$ ($X_1^-$). The discontinuities in the PL energy between regions I/II and I/III are attributed to CIE binding energies of ~ 15 meV and ~ 10 meV, respectively[19]. In region IV, when the electron density is further increased, an additional PL peak, $X_2^-$, appears ~ 25 meV above the $X_1^-$ peak, which overtakes in intensity with increasing $n_{2D}$. This additional exciton feature is likely related to reaching the upper conduction band of MoSe$_2$ (CB2). For device B (Fig. 1f), we observe a similar discontinuity in emission energy when entering regions II and III due to CIE formation but find the higher energy peak to always be present as we tune the carrier density.

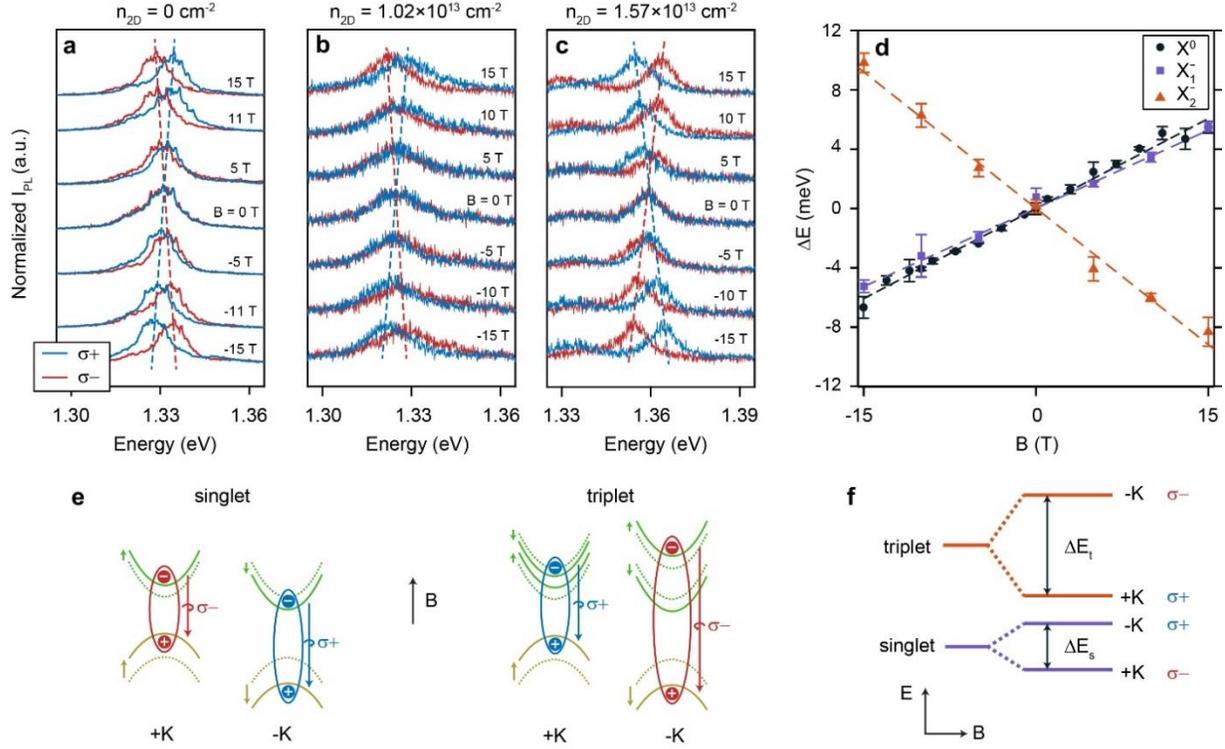

**Figure 2 | Experimental evidence of spin-singlet and spin-triplet excitons via magnetic field in device A. a-c,** Polarization-resolved photoluminescence (PL) spectra at $n_{2D}$ = 0, 1.02 × 10$^{13}$ cm$^{-2}$, and 1.57 × 10$^{13}$ cm$^{-2}$, respectively, for characteristic magnetic fields (*B*) using a cross-polarized measurement scheme (Supplementary Section 4). Blue (red) curves are σ+ (σ−) collection. The dashed lines serve as a guide to the eye. **d,** PL energy splitting ($\Delta E = E_{\sigma+} - E_{\sigma-}$) as a function magnetic field for $X^0$, $X_1^-$, and $X_2^-$. Error bars are calculated from the fitting of the peak position. The dashed lines are linear fits to the energy splitting giving $g_0 = 6.99 \pm 0.35$, $g_1 = 6.06 \pm 0.58$, and $g_2 = -10.6 \pm 1.0$. **e,** Band diagrams and Zeeman splitting for the singlet and triplet transitions without (dashed lines) and with (solid lines) magnetic field. **f,** Zeeman splitting for the singlet and triplet excitons in the exciton particle picture and the circular polarization light coupling. The triplet exciton has an enhanced *g*-factor due to spin contributions.

To understand the angular momentum characteristics of the interlayer excitons, we measure PL under magnetic fields to determine the effective Zeeman splitting of the exciton species. We perform polarization-resolved PL measurements as a function of magnetic field (*B*) using a cross-polarized measurement scheme (Supplementary Fig. 4a). Figs. 2a-c show the normalized σ+ (blue) and σ− (red) PL spectra measured in device A at $n_{2D}$ = 0, 1.02 × 10$^{13}$ cm$^{-2}$, and 1.57 × 10$^{13}$ cm$^{-2}$, respectively. From these polarization-resolved spectra, we obtain the PL energy splitting between the circularly polarized light ($\Delta E = E_{\sigma+} - E_{\sigma-}$) as a function of *B*. Fig. 2d shows the measured energy difference follows a linear relation $\Delta E = g\mu_B B$, where *g* is the effective *g*-factor and $\mu_B$ is the Bohr magneton. From the slope of the measured relation between $\Delta E$ and *B*, we obtain the effective *g*-factors for $X^0$, $X_1^-$, and $X_2^-$: $g_0 = 6.99 \pm 0.35$, $g_1 = 6.06 \pm 0.58$, and $g_2 = -10.6 \pm 1.0$, respectively. Interestingly, the *g*-factor for $X_2^-$ is greater than and has the opposite sign of $g_0$ and $g_1$, implying $X_2^-$ has an additional Zeeman splitting contribution and that the chiral light coupling to the K valleys is flipped compared to $X^0$ or $X_1^-$. The observation of a higher energy emission in region IV also suggests that transitions between the highest WSe$_2$ K-valley valence band and both spin-split MoSe$_2$ K-valley conduction bands are allowed. This would indicate that the higher energy peak is an emissive triplet transition with an in-plane dipole moment, unlike dark triplet excitons in monolayers[29–31].

The experimental observations of ~ 25 meV splitting between $X_2^-$ and $X_1^-$ as well as the opposite polarization characteristics are consistent with spin-singlet and -triplet transition selection rules in the 0-degree aligned heterostructure (Supplementary Section 5). Quantitative evidence for the singlet and triplet states and opposite circular polarization coupling is revealed by calculating the expected exciton g-factors using a single electron band picture[17,32]. The expected g-factor is based on the Zeeman shift of each electron band (Fig. 2e) without considering any additional excitonic effects under magnetic field (further details in Supplementary Section 6). From this model, we calculate the singlet and triplet g-factors to be $g_{singlet}^{0-theory} \approx 7.1$ and $g_{triplet}^{0-theory} \approx -11.1$, respectively. These calculated g-factors are in excellent agreement with experimentally observed values both in terms of sign and magnitude. Thus, we confirm $X_1^-$ and $X_2^-$ as singlet and triplet excitonic transitions, respectively. We note that unlike the traditional picture of singlet and triplet states, the degeneracy of interlayer exciton singlet and triplet states is already broken due to spin-orbit coupling. These states split differently under magnetic field as shown in Fig. 2f in the exciton particle picture. Similarly, we performed magneto-PL measurement in a separate 60-degree aligned sample and confirmed the selection rules and associated large triplet transition g-factor, $g_{triplet}^{60-stack} \approx -15.1 \pm 0.4$ (Supplementary Section 7) in agreement with theory and previous measurements[28]. Thus, we confirm the emission of triplet excitons in 0-degree heterostructures occur, but only at high $n_{2D}$.

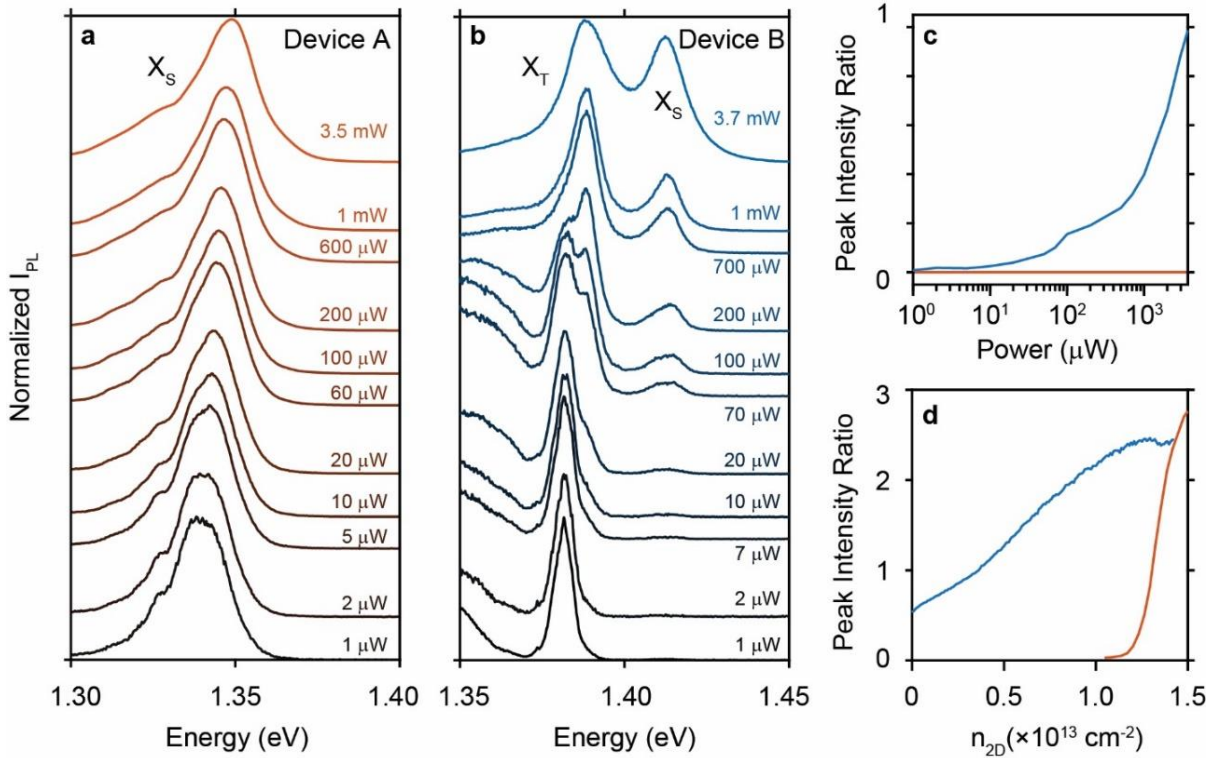

**Figure 3 | Tunable interlayer exciton species via excitation power and carrier density. a, b,** Power (*P*) dependence of the normalized photoluminescence (PL) spectra at $V_{tg} = \alpha\, V_{bg} = 0$ V for devices A and B, respectively. **c,** Peak intensity ratio between the higher energy peak and the lower energy peak for device A (orange) and device B (blue) as a function of excitation power at $V_{tg} = \alpha\, V_{bg} = 0$ V. The higher energy peak does not appear at this carrier density in device A. **d,** The same peak intensity ratio as a function of carrier density at *P* = 6 µW for device A (orange) and *P* = 100 µW for device B (blue).

From this analysis, we can now assign the peaks in the PL spectra as either singlet or triplet states. In regions I-III, $X^0$, $X^+$, and $X_1^-$ all have transitions from CB1, allowing us to assign them as singlet neutral or singlet charged excitons. In region IV, the $X_2^-$ peak is a transition from CB2 in the presence of free carriers and is therefore a triplet charged exciton. Energetically, the triplet charged exciton can form with an electron in CB1 of either K valley in MoSe$_2$, but further studies are required for a more detailed understanding. The emergence of $X_2^-$ only after sufficient band filling can be explained by the relative dipole strengths of the singlet and triplet excitons. The lifetimes of $X_1^-$ and $X_2^-$ at high $n_{2D}$ were measured to be $\tau_1 = 6.08 \pm 0.01$ ns and $\tau_2 = 6.12 \pm 0.02$ ns (Supplementary Fig. 8 inset), respectively, suggesting the optical dipole strength of the two exciton species are similar, consistent with theoretical calculations[5].

We find that the relative intensity of singlet and triplet exciton emission can be tuned by electrostatic doping and power of the laser excitation. Figs. 3a-b show the power dependence of the normalized PL emission in devices A and B. The double peaked features near 1.34 eV and 1.37 eV in devices A and B, respectively, are attributed to CIEs due to changes in residual doping by the excitation power. We find that device A at neutral doping is always dominated by lower energy singlet emission without triplet emission even at high excitation powers. Device B even at neutrality, on the other hand, has a higher energy singlet peak (~ 1.42 eV) which becomes more prominent at higher powers. In the strongly non-equilibrium state at the largest excitation, the singlet and triplet emission are about similar intensity (Fig 3c). While the excitation power can show limited range of singlet/triplet emission ratio, the electrostatic doping tuned by the gate can vary the triplet/singlet emission ratio in a wide range. Fig. 3d shows that initially dominant singlet emission in device A turned to more than 70% triplet emission in the high doping range ($n_{2D} > 1.4 \times 10^{13}$ cm$^{-2}$). In device B, the inverse ratio can be tuned, with the singlet/triplet emission ratio varying between 0.5 – 2.5 over a similar carrier density range. Thus, the MoSe$_2$/WSe$_2$ heterostructure provides a platform for fully tunable singlet-triplet exciton emission depending via electrical gates, excitation power, and the stacking registry.

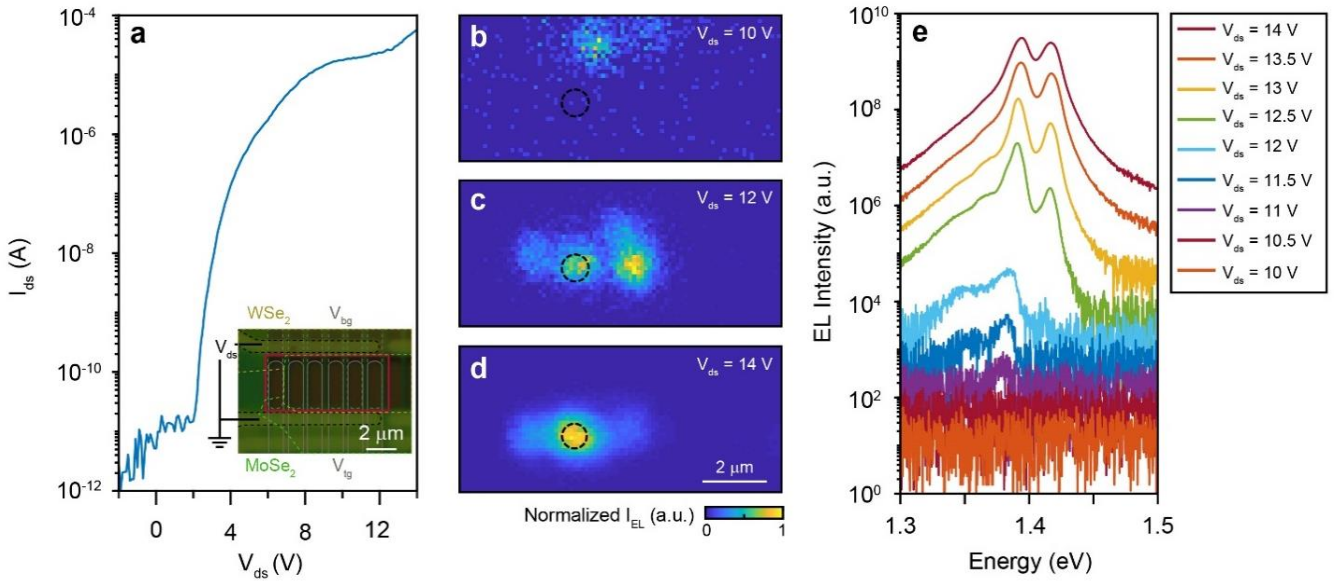

**Figure 4 | Electrical generation of singlet and triplet exciton species in device B. a,** I-V characteristics of the device B at $V_{tg} = \alpha\, V_{bg} = 10$ V, where α = 1.4 is based on the h-BN thicknesses. Inset: Optical image of the sample with yellow (green) dashed line indicating WSe$_2$ (MoSe$_2$) area, gray solid (dashed) lines indicating $V_{tg}$ ($V_{bg}$), black dashed lines outline the contacts, and the red rectangle to indicate sample area of interest in (b-d). **b-d,** Spatial maps of normalized electroluminescence (EL) generated from the sample when in forward bias ($V_{ds}$ = 10, 12 and 14 V). The black dashed circle indicates the collection spot for (e). **e,** EL spectra as a function of $V_{ds}$ at $V_{tg} = \alpha\, V_{bg} = 10$ V. Each curve is vertically offset by $10^{0.5}$.

Utilizing electrical contacts in WSe$_2$ and MoSe$_2$ layers, we can operate the device as a gate tunable atomically thin PN diode[15], where the interlayer tunneling current across the vdW interface can generate singlet and triplet exciton emission. Fig. 4a shows the current ($I_{ds}$) vs. drain-source voltage ($V_{ds}$) curve at $V_{tg} = α\ V_{bg} = 10$ V, where $α = 1.4$, demonstrating rectifying diode behavior as expected for a type-II aligned heterostructure. We select this gate voltage to ensure emission in the heterostructure region under high bias (see Supplementary Section 9). Figs. 4b-d show the spatial distribution of the electroluminescence (EL) emission from the red outlined area of the heterostructure (Fig. 4a inset) under the same gate conditions at different $V_{ds}$ in the high bias regime. We find that the spatial distribution of the emission is inhomogeneous, presumably due to disorders in the channel and the lateral gaps between the gate structures (Supplementary Section 9). The emission position shifts sensitively with $V_{ds}$ and $V_{tg}$, which tune the current distribution in the channel (see Supplementary Section 10). Fig. 4e shows the EL spectra collected at a fixed location of the sample (marked by dashed black circle in Figs. 4b-d). With sufficient $V_{ds}$, we observe EL emission from both the singlet and triplet exciton peaks. We note that for $V_{ds} < 12$ V, the emission spot moves away from the collection spot, explaining the lack of emission signal. Regardless, the EL of singlet and triplet excitons demonstrates complete electrical generation of excitonic states with opposite spin and polarizability characteristics. The large energy splitting provides sufficient energy separation to allow coupling to the two excitonic species independently from each other. Furthermore, we observe a similar amount of exciton emission from both states, effectively doubling the light emission in comparison to systems with disallowed triplet exciton transitions.

Our capability of gate tuning to access the higher conduction band with opposite spin allows us to create charged excitons with singlet and triplet spin configurations and opposite chiral light coupling. Electrical generation of tunable singlet and triplet excitons in vdW heterostructures, combining long EL lifetime[19] with local gate engineering[33], paves the way towards independently controlling chiral, valley, and spin quantum states in valleytronic devices.

**References**

1. Rivera, P. *et al.* Valley-polarized exciton dynamics in a 2D semiconductor heterostructure. *Science.* **351**, 688–691 (2016).
2. Hsu, W. T. *et al.* Negative circular polarization emissions from $WSe_2/MoSe_2$ commensurate heterobilayers. *Nat. Commun.* **9**, 1356 (2018).
3. Ciarrocchi, A. *et al.* Polarization switching and electrical control of interlayer excitons in two-dimensional van der Waals heterostructures. *Nat. Photonics* **13**, 131–136 (2019).
4. Hanbicki, A. T. *et al.* Double Indirect Interlayer Exciton in a $MoSe_2/WSe_2$ van der Waals Heterostructure. *ACS Nano* **12**, 4719–4726 (2018).
5. Yu, H., Liu, G. Bin & Yao, W. Brightened spin-triplet interlayer excitons and optical selection rules in van der Waals heterobilayers. *2D Mater.* **5**, 035021 (2018).
6. Wu, F., Lovorn, T. & Macdonald, A. H. Theory of optical absorption by interlayer excitons in transition metal dichalcogenide heterobilayers. *Phys. Rev. B* **97**, 035306 (2018).
7. Kido, J., Kimura, M. & Nagai, K. Multilayer white light-emitting organic electroluminescent device. *Science.* **267**, 1332–1334 (1995).
8. Becker, M. A. *et al.* Bright triplet excitons in caesium lead halide perovskites. *Nature* **553**, 189–193 (2018).
9. Mak, K. F., Lee, C., Hone, J., Shan, J. & Heinz, T. F. Atomically Thin $MoS_2$: A New Direct-Gap Semiconductor. *Phys. Rev. Lett.* **105**, 136805 (2010).
10. Scuri, G. *et al.* Large Excitonic Reflectivity of Monolayer $MoSe_2$ Encapsulated in Hexagonal Boron Nitride. *Phys. Rev. Lett.* **120**, 37402 (2018).
11. Fang, H. H. *et al.* Control of the Exciton Radiative Lifetime in van der Waals Heterostructures. *Phys. Rev. Lett.* **123**, 1–14 (2019).
12. Chernikov, A. *et al.* Exciton binding energy and nonhydrogenic Rydberg series in monolayer WS2. *Phys. Rev. Lett.* **113**, 076802 (2014).
13. Mak, K. F., He, K., Shan, J. & Heinz, T. F. Control of valley polarization in monolayer $MoS_2$ by optical helicity. *Nat. Nanotechnol.* **7**, 494–498 (2012).
14. Onga, M., Zhang, Y., Ideue, T. & Iwasa, Y. Exciton Hall effect in monolayer $MoS_2$. *Nat. Mater.* **16**, 1193–1197 (2017).
15. Lee, C. H. *et al.* Atomically thin p-n junctions with van der Waals heterointerfaces. *Nat. Nanotechnol.* **9**, 676–681 (2014).
16. Baugher, B. W. H., Churchill, H. O. H., Yang, Y. & Jarillo-Herrero, P. Optoelectronic devices based on electrically tunable p-n diodes in a monolayer dichalcogenide. *Nat. Nanotechnol.* **9**, 262–7 (2014).
17. Nagler, P. *et al.* Giant magnetic splitting inducing near-unity valley polarization in van der Waals heterostructures. *Nat. Commun.* **8**, 1551 (2017).
18. Miller, B. *et al.* Long-Lived Direct and Indirect Interlayer Excitons in van der Waals Heterostructures. *Nano Lett.* **17**, 5229–5237 (2017).
19. Jauregui, L. A. *et al.* Electrical control of interlayer exciton dynamics in atomically thin heterostructures. *Science.* **366**, 870–875 (2019).
20. Rivera, P. *et al.* Observation of long-lived interlayer excitons in monolayer $MoSe_2$–$WSe_2$ heterostructures. *Nat. Commun.* **6**, 6242 (2015).
21. Wang, J. *et al.* Optical generation of high carrier densities in 2D semiconductor heterobilayers. *Sci. Adv.* **5**, eaax0145 (2019).
22. Ceballos, F., Bellus, M. Z., Chiu, H. Y. & Zhao, H. Ultrafast charge separation and indirect exciton formation in a $MoS_2$-$MoSe_2$ van der waals heterostructure. *ACS Nano* **8**, 12717–12724 (2014).
23. Chen, H. *et al.* Ultrafast formation of interlayer hot excitons in atomically thin $MoS_2/WS_2$ heterostructures. *Nat. Commun.* **7**, 12512 (2016).

24. Hong, X. *et al.* Ultrafast charge transfer in atomically thin MoS$_2$/WS$_2$ heterostructures. *Nat. Nanotechnol.* **9**, 682–686 (2014).
25. Kunstmann, J. *et al.* Momentum-space indirect interlayer excitons in transition-metal dichalcogenide van der Waals heterostructures. *Nature Physics* vol. 14 801–805 (2018).
26. Wilson, N. R. *et al.* Determination of band offsets, hybridization, and exciton binding in 2D semiconductor heterostructures. *Sci. Adv.* **3**, e1601832 (2017).
27. Zhang, L. *et al.* Highly valley-polarized singlet and triplet interlayer excitons in van der Waals heterostructure. *Phys. Rev. B* **100**, 1–14 (2019).
28. Wang, T. *et al.* Giant Valley-Zeeman Splitting from Spin-Singlet and Spin-Triplet Interlayer Excitons in WSe$_2$/MoSe$_2$ Heterostructure. *Nano Lett.* (2019) doi:10.1021/acs.nanolett.9b04528.
29. Zhou, Y. *et al.* Probing dark excitons in atomically thin semiconductors via near-field coupling to surface plasmon polaritons. *Nat. Nanotechnol.* **12**, 856–860 (2017).
30. Zhang, X. X., You, Y., Zhao, S. Y. F. & Heinz, T. F. Experimental Evidence for Dark Excitons in Monolayer WSe$_2$. *Phys. Rev. Lett.* **115**, 257403 (2015).
31. Echeverry, J. P., Urbaszek, B., Amand, T., Marie, X. & Gerber, I. C. Splitting between bright and dark excitons in transition metal dichalcogenide monolayers. *Phys. Rev. B* **93**, 121107(R) (2016).
32. Wang, G. *et al.* Magneto-optics in transition metal diselenide monolayers. *2D Mater.* **2**, 034002 (2015).
33. Wang, K. *et al.* Electrical control of charged carriers and excitons in atomically thin materials. *Nat. Nanotechnol.* **13**, 128–132 (2018).